\documentclass[10pt,article,oneside]{memoir}
\pdfoutput=1

\date{}
\counterwithout{section}{chapter}

\usepackage{amsthm, amssymb, amsmath, latexsym, amsfonts, mathcomp}
\usepackage{xfrac}
\usepackage{verbatim}
\usepackage{color}
\usepackage{mathrsfs}
\usepackage{listings}
\usepackage{array}
\usepackage{url}
\usepackage{footnote}
\usepackage[style=authortitle-ibid,url=true,doi=false,isbn=false]{biblatex}
\usepackage{microtype}

\bibliography{Everything}
\usepackage{graphicx}

\allowdisplaybreaks

\newtheorem{proposition}{Proposition}[chapter]
\newtheorem{theorem}{Theorem}[chapter]
\newtheorem{lemma}{Lemma}[chapter]
\newtheorem{corollary}{Corollary}[chapter]
\newtheorem{definition}{Definition}[chapter]
\newtheorem{postulate}{Postulate}[chapter]
\newtheorem{axiom}{Axiom}[chapter]

\theoremstyle{definition}
\newtheorem{remark}{Remark}[chapter]

\def\dd{\mathrm{d}}
\def\Den{\mathrm{Den}}

\def\R{\mathbb{R}}

\def\M{\mathcal{M}}

\def\Lie{\pounds}
\newcommand{\dpp}[2]{\ensuremath{\frac{\partial #1}{\partial #2}}}

\DeclareCiteCommand{\parencite}[\mkbibparens]
  {\usebibmacro{prenote}}
  {\usebibmacro{citeindex}%
    \clearfield{url}%
    \usebibmacro{cite}}
  {\multicitedelim}
  {\usebibmacro{cite:postnote}}
\DeclareCiteCommand{\footcite}[\mkbibfootnote]
  {\usebibmacro{prenote}}
  {\usebibmacro{citeindex}%
    \clearfield{url}%
    \usebibmacro{cite}}
  {\multicitedelim}
  {\usebibmacro{cite:postnote}}

\DeclareCiteCommand{\parencitenn}[\mkbibparens]
  {\usebibmacro{prenote}}
  {\usebibmacro{citeindex}%
   \usebibmacro{citetitle}}
  {\multicitedelim}
  {\usebibmacro{cite:postnote}}

\DeclareCiteCommand{\citenn}
  {\usebibmacro{prenote}}
  {\usebibmacro{citeindex}%
   \usebibmacro{citetitle}}
  {\multicitedelim}
  {\usebibmacro{cite:postnote}}

\title{On the Mathematical Formulation of Radiance}

\author{\small{Christian Lessig$^{1,2}$ and Eugene Fiume$^1$ and Mathieu Desbrun$^2$} 
        \\
        \small{$^1$ Dynamic Graphics Project, University of Toronto}
        \\
        \small{$^2$ Computational + Mathematical Sciences, California Institute of Technology}
}

\begin{document}

\maketitle

\vspace{-0.35in}
\begin{abstract} 

Radiance is widely regarded as the principal quantity in light transport theory.
Yet, the concept of radiance in use today has remained mostly unchanged since Lambert's work in the 18$^\textrm{th}$ century. His formulation of the measurement of light intensity is based on classical differentials, and is known to suffer from several theoretical and practical limitations. 
After tracing the historic development of radiance and its shortcomings, we provide a modern formulation of light intensity measurements that models radiance as a differential form. We demonstrate the utility of this use of exterior calculus 
for questions in light transport theory by rigorously deriving the cosine term and the area formulation, without the need for postulates or heuristic arguments. 
% as an effective language to study 
%We also discuss how this formulation of radiance fits within a modern treatment of a general theory of light transport.
The formulation of radiance as a differential form introduced in this paper hence provides the first step towards a modern theory of light transport.
\end{abstract}

\section{A Short History of Radiometry}
\label{sec:radianc:history}

In ancient times, light intensity was believed to emanate from the eye as \emph{fire}, one of the four elements.
Already in the Middle Ages, however, it was realized that the intensity is an objective quantity that travels along light rays towards the eye.
%along light rays towards the eye.
The first notable contributions toward a quantitative understanding of light intensity and its measurement were made in the 17$^{\textrm{th}}$ century by Kepler and Mersenne who established the inverse square law and the cosine law for incident radiation.
%, respectively.
In the 18$^{\textrm{th}}$ century, the work of Bouguer\footcite{Bouguer1729,Bouguer1760} and Lambert\footcite{Lambert1760} founded radiometry as a research field in its own right.
%studied independently of geometrical optics.
Bouguer's work was concerned for example with the decay of light intensity in the atmosphere, which led him to what is now known as Beer-Lambert law.
He also studied the scattering of light on rough surfaces modelled by specular microfacets.
Lambert's central contributions in his \emph{Photometrica} were:
\begin{itemize}
  \item the systematization of radiometry based on postulates firmly rooted in experiments;
  \item and the introduction of infinitesimals to quantitatively describe the measurement of light intensity.
\end{itemize}
To model the measurement of light flux at a surface point in a direction, Lambert also introduced \emph{splendor}, corresponding to luminance---the photometric sibling of radiance---in today's nomenclature\footcite{Lambert2001}.
The physics of light played no role for Lambert, and in environments without absorption he employed an ``actio in distans'' principle to related outgoing to incoming intensity.\footcite{Gershun1936,DiLaura2011}
Beginning in the 18$^{\textrm{th}}$ century, the relevance of radiometry for applications began to increase, and for example Euler employed it to study astronomical questions,\footcite{Euler1752} Lagrange in work on optical systems,\footcite{Lagrange1803} and Fourier in his seminal treatment of heat transfer.\footcite{Fourier1822,Fourier1825}

The use of radiometry to study questions in optics,\footcite{Helmholtz1867,Helmholtz1874} astronomy,\footcite{Herschel1828,Zoellner1865} and heat transport,\footcite{Kirchhoff1859,Clausius1864,Planck1912} continued throughout the 19$^{\textrm{th}}$ century.
%the latter gaining particular prominence through the rise of thermodynamics, 
Usually, however, the work employed only ``the well-known laws of photometry''\footcite{Helmholtz1874} and even Lambert's \emph{Photometrica} was rarely mentioned.
Consequently, many important results were rediscovered over time.
For example, the quadratic dependence of classical radiance on the refractive index was obtained independently for heat\footcite{Kirchhoff1859} and optical radiation\footcite{Herzberger1928}. 
Although many prominent scientists and distinguished mathematicians, such as von Helmholtz\footcite{Helmholtz1867,Helmholtz1874}, Kirchhoff\footcite{Kirchhoff1859}, and Clausius\footcite{Clausius1864}, employed radiometry, the physical and mathematical foundations of the theory remained entrenched in the $18^{\textrm{th}}$ century.
A notable exception is the work by Lommel\footcite{Lommel1880,Lommel1889}.
He considered radiance as emanating from a volume, an idea that was forgotten shortly afterwards, and also extended the Beer-Lambert law to include out-scattering.
Similar results that accounted for scattering were obtained by Chwolson\footcite{Chwolson1889} and somewhat later by Schuster,\footcite{Schuster1905} while Schwarzschild\footcite{Schwarzschild1906} studied radiation equilibrium, a question again considered before by Lommel\footcite{Lommel1889}
Based on Schuster's work,\footcite{Schuster1905} Jackson\footcite{Jackson1910} and King\footcite{King1913} developed the transport equations that are in use to this date. 
The classical theory of radiometry culminated in the work by Nicodemus.\footcite{Nicodemus1963,Nicodemus1977z}
In contrast to Lambert's postulates, he employed heuristic geometric arguments to establish the fundamental laws of radiometry.
%, although without the careful experiments that justified the results in \emph{Photometrica}.
Nonetheless, conceptually Nicodemus' work belongs to the tradition of the 18$^{\textrm{th}}$ and 19$^{\textrm{th}}$ century.
%In contrast to Lambert's postulates, however, Nicodemus employed heuristic geometric arguments,
%For heat radiation, the culmination point of the classical theory of radiometry can be found in the booklet by Planck~\cite{Planck1912}, that is still cited to this date, cf.~\cite{Wolf1978}.

In the first half of the 20$^\textrm{th}$ century, the introduction of methods from vector calculus led to a considerable improvement in the mathematical formulation of radiometric theory\footcite{Gershun1936,Moon1953,Moon1981}.
Vector irradiance led to closed form solutions for the irradiance for many geometric configurations\footcite{Gershun1936,Fock1924,Schroder1993} and Arvo also employed the concept for light transport simulation.\footcite{Arvo1994}
As already recognised in 1939, a formulation of radiometry based on vector calculus is however limited since: ``[v]aluable as the methods [using vector calculus] are, however, they probably do not constitute the ultimate solution of the problem. [\ldots] the physically important quantity is actually the illumination, which is a function of five independent, not three.''.\footcite[p. 51]{Gershun1936}
In the second half of the 20$^\textrm{th}$ century, the physical foundations of radiometry and light transport were finally reconsidered.
Early work\footcite{Duderstadt1979,Arvo1993a} was based on particle models.
This approach suffered however from an interpretation of light particles as photons which is physically not meaningful, cf. also the work by Mishchenko.\footcite{Mishchenko2011}
Later approaches\footcite{Wolf1978,Kim1987,Ryzhik1996,Bal2006,Mishchenko2006a,Mishchenko2007} were physically more earnest.
However, these efforts were limited by misconceptions about the nature of radiance, for example that it can only be expected to be meaningful in macroscopic environments\footnote{See for example~\parencite{Wolf1978,Kim1987}.} or that it represents an infinitesimal surface flux.\footnote{Cf.~\parencite{Bal2006}.}
%Additionally, these recent developments have so far not been adopted or recognized by practitioners.
The 20$^{\textrm{th}}$ century saw also many more applications of radiometry, for example in medical imaging, remote sensing, atmospheric science, astrophysics, and, last but not least, in computer graphics.
However, the theoretical developments of the $20^\textrm{th}$ century were only rarely adopted by practitioners and most often the classical theory from Lambert's time was and is employed.

\section{Radiometry Reconsidered}

As we have seen, in the past 250 years radiometry has found applications in a wide range of fields---while most of the time it was not considered a subject worthy to be studied in its own right.
The radiometric theory employed in practice is hence still that developed by Lambert in the 1750s.
Central to this classical theory is the radiance
\begin{align}
  L(x,\omega) \, \cos{\theta}  \, d\omega \, dA 
\end{align}
which represents the infinitesimal flux of light energy in direction $\omega \in S^2$ through a surface with normal $\vec{n}$, with $\theta$ being the angle between $\omega$ and $\vec{n}$.\footcite{Dutre2006,pbrt2}
The importance of radiance lies in the ability to derive all other radiometric quantities---and the observable finite energy flux---from it by integration. 
%the lack of reconsideration of its physical and mathematical foundations in the past 250 years, 
Despite this crucial role, however, the current conceptualization of radiance suffers from two central limitations:
\begin{enumerate}
  \item classical infinitesimals are employed;
  \item radiance is employed to model transport, although its mathematical formulation represents surface flux.
\end{enumerate}
As long as 80 years ago, these shortcomings led to a characterization of radiance as ``absurdly antiquated''\footcite[p. 51]{Gershun1936} and in fact it is the current formulation that makes it necessary to introduce the cosine law and the area formulation as postulates or based on qualitative geometrical arguments, such as those commonly employed in the current computer graphics literature.\footcite{Dutre2006,pbrt2}

The current formulation of radiance using classical infinitesimals and its misuse to describe transport has led to ample theoretical and practical limitations.
For example, for volume transport, where an ``actio in distans'' principle is not applicable, the length along the ray must be introduced a posteriori to obtain a volumetric description that can account for absorption or volume scattering.
Likewise, in media with varying refractive index, an approach based on area fluxes is artificial.
Because of this, the quadratic dependence of radiance on the refractive index can only be obtained using ad-hoc derivations such as in.\footcite{Veach_phd}
Moreover, to describe transport in such inhomogeneous media, concepts from geometrical optics and radiometry have to be blended, even when no connection between them has been rigorously established to this date.\footnote{Cf. for example~\parencite{Ihrke2007,Zhao2007}.}
Another significant problem with the current understanding of radiance is that differentiation along a ray is not possible.
This prevents the use of derivative values in ray tracing, which would provide an avenue to better amortize the prohibitive costs of tracing rays by obtaining two or more values of the light energy signal from each single ray.
For techniques that interpolate light energy, and which are frequently employed in practical applications, the intrinsic character of the light energy has to be respected: when interpolation is to be performed in space, such as in photon mapping, a volumetric formulation of radiance is required, while on surfaces, interpolation must ensure that measurements are correctly reproduced.

%Despite of the central place of light transport theory in computer graphics, solid foundations for the theory are still missing.
%, leading to numerous theoretical and practical problems.
In the present paper, we introduce a modern mathematical conceptualization of radiance that overcomes the limitations associated with the measurement of light intensity in classical radiometry.
Our formulation of radiance as a differential form enables us to employ the geometric language of exterior calculus to rigorously establish the well-known cosine term and the area formulation, without the need for postulates or heuristic arguments as currently employed in the literature.
Additionally, much insight into the nature of radiance is obtained.
To avoid most of the complexity of modern light transport theory,\footcite{Lessig_phd} we restrict the following exposition to radiance and questions related to measurements. 
A comprehensive discussion, including a rigorous physical justification for the formulation of radiance we provide here, will be presented in a forthcoming publication.
Some intuition is already provided in the appendix, and the intuition behind differential forms is provided in

\section{A Modern Formulation of Radiance}
\label{sec:radiance:modern}

Radiance describes the infinitesimal light energy flowing through a surface $S \subset \R^3$ from direction $\bar{p} \in S^2$ in unit time and unit frequency.
A finite measurement of energy flux is thus obtained by integration, which, to be physically meaningful, has to be covariant, e.g., independent of the coordinate system.
Such covariant ``integrands'' are naturally modelled by differential forms--=a contemporary mathematization of classical infinitesimals.
%a differential form, both for the surface integral and the integral over the space of all directions.
%For the energy flow to be covariant, that is independent of the coordinate system that is employed, and well defined, radiance is required to be a differential form.
%Moreover, because radiance is to be integrated over a surface, it has to be a differential $2$-form, and since it is also to be integrated over the space of directions, it also has to be a differential form over this space.
A modern formulation of radiance can thus be written as
\begin{align}
  \Lambda = L(q,\bar{p},\nu,t) \, dA_{\perp} \, d\bar{p} \, d\nu \, dt
  \label{eq:radiance:modern:1}
\end{align}
where $\nu$ and $t$ are frequency and time, respectively, $\bar{p} = (\bar{p}_1,\bar{p}_2,\bar{p}_3)$ is a unit direction (co-)vector and $d\bar{p}$ the infinitesimal solid angle form in this direction.
The differential $2$-form $dA_{\perp} \in \Omega^2(\R^3)$ is given by
\begin{align}
  dA_{\perp}(\bar{p}) = \bar{p}_1 \, dy \wedge dz + \bar{p}_2 \, dz \wedge dx + \bar{p}_3 \, dx \wedge dy .
  \label{eq:radiance:modern:area_2_form}
\end{align}
In contrast to the classical scalar-valued radiance, our formulation $\Lambda$ of the concept is vector-valued, as is more apparent when written as
\begin{align}
  \Lambda =
  ( \quad & L(q,\bar{p},\nu) \, \bar{p}_1 \, dy \wedge dz
  \nonumber \\
  + & \, L(q,\bar{p},\nu) \, \bar{p}_2 \, dz \wedge dx
  \label{eq:radiance:modern:2} \\
  + & \, L(q,\bar{p},\nu) \, \bar{p}_3 \, dx \wedge dy ) \, d\bar{p} \, d\nu
  \nonumber
\end{align}
or, when one omits the basis functions and just writes the components, as
\begin{align}
  \Lambda = (L(q,\bar{p},\nu) \bar{p}_1, L(q,\bar{p},\nu) \bar{p}_2, L(q,\bar{p},\nu) \, \bar{p}_3 ) .
  \label{eq:radiance:modern:components}
\end{align}
As will be seen in the next section, $\Lambda$ becomes scalar only when pulled back to the chart of a surface, and it is this pullback that naturally leads to the cosine term.

%\begin{figure}
%  \frame{
%  \begin{minipage}{1.0\columnwidth}
%    \setlength{\leftskip}{0.05in}
%    \setlength{\rightskip}{0.05in}
%    \vspace{0.05in}
%    {\bfseries What is the differential $2$-form $dA_{\perp}$?}
\begin{remark}
  {\itshape
    The most significant difference between our conceptualization of radiance and the classical one is the use of the $2$-form $dA_{\perp}$ in Eq.~\ref{eq:radiance:modern:area_2_form}.
    A $2$-form represents flux through an arbitrary surface as the linear superposition of fluxes through elementary $2$-forms $dy \wedge dz$, $dz \wedge dx$, and $dx \wedge dy$ aligned with the coordinate axes.
    These elementary $2$-forms are the $2$-form basis vectors which represent the flux along the $e_x$, $e_y$, and $e_z$ axis, respectively, see the figure below.
    \begin{center}
    \includegraphics[trim = 70mm 185mm 62mm 23mm, clip, scale=0.8]{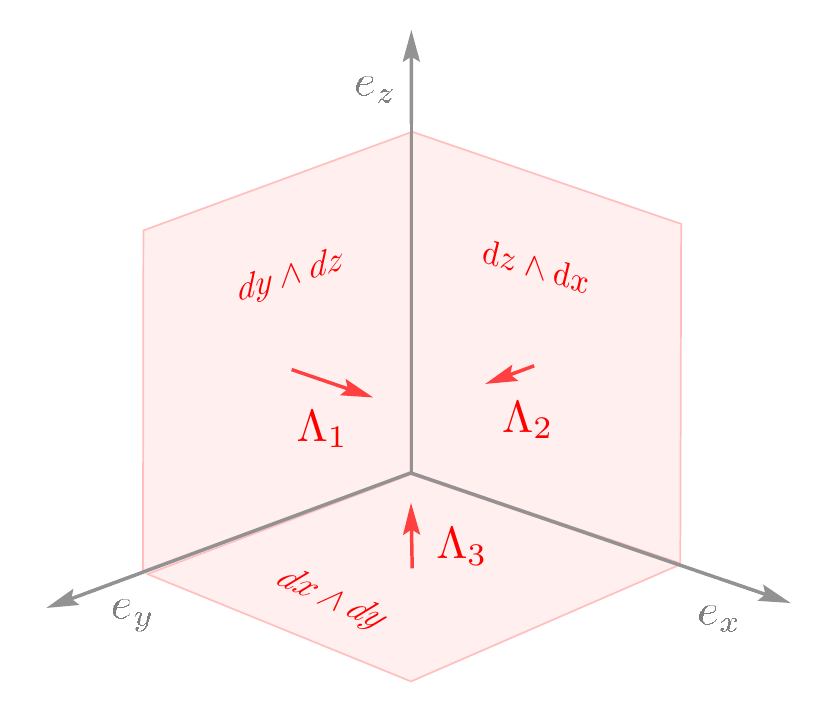}
  \end{center}
  Radiance differs from a classical, scalar $2$-form in that it also has an angular dependence on $\bar{p}$, and these two dimensions agree in that $\bar{p}$ represents the direction of maximal flux, or the surface orientation for which the flux represented by $dA_{\perp}$ is maximized.
  }
\end{remark}
%  \vspace{0.05in}
%  \end{minipage}
%  }
%  \label{fig:radiance:two_form}
%  \vspace{-0.2in}
%\end{figure}

Although vector-valued, our formulation of radiance should not be confused with vector irradiance $\mathrm{J}$ which is related to radiance through the integral
\begin{align}
  \mathrm{J} = \left( \int_{H_q^2} L(q,\bar{p},\nu) \, d\bar{p} \, d\nu \right) dA_{\perp} \, d\nu \, dt ,
  \label{eq:vector_irradiance}
\end{align}
where ${H_q^2}$ is the positive hemisphere over $q \in \R^3$ as defined by $\bar{p}$.
Eq.~\ref{eq:vector_irradiance} implies that vector irradiance is also a differential $2$-form, that is
\begin{align}
  \mathrm{J} = J(q) \, dA_{\perp} \, d\nu \, dt ,
\end{align}
a fact which is also apparent when one re-reads Gershun's paper\footcite[Chapter IV.1]{Gershun1936} with the modern theory in mind.
The difference between vector irradiance and radiance is hence that the former is a ``pure'' $2$-form, that is $\mathrm{J} \in \Omega^2(\R^3)$, whereas for the latter we have $\Lambda \in \Omega^2(\R^3) \otimes \Omega^2(T_q^*Q)$, omitting in both cases the frequency and time dimension for clarity.

In the following, we derive the cosine term and the area formulation directly from our formulation of radiance using the modern language of exterior calculus of differential forms.

%%%%%%%%%%%%%%%%%%%%%%%%%%%%%%%%%%%%%%%%%%%%%%%%%%%%%%%%%%%%%%%%%%%%%%%%%%%%%%%%
\subsection{The Cosine Term}
\label{sec:radiance:cosine}

\begin{figure*}[t]
  \setlength{\abovecaptionskip}{-10pt}
  \begin{center}
    \includegraphics[trim = 38mm 175mm 30mm 25mm, clip, scale=0.8]{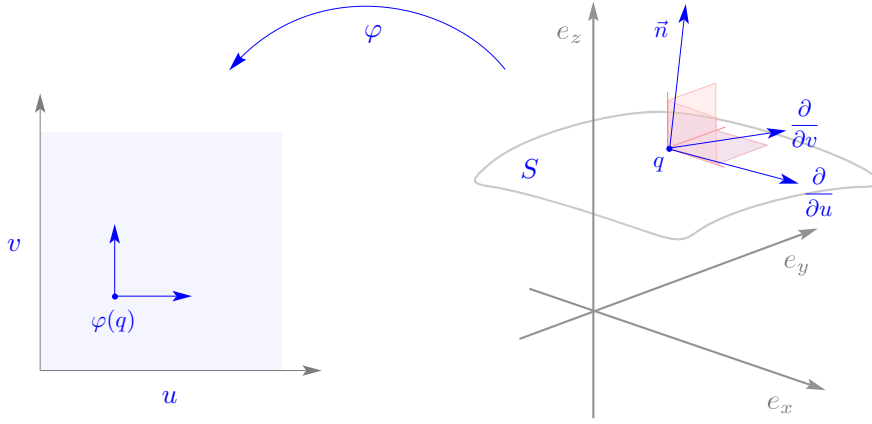}
\end{center}
  \caption{Geometry of the pullback of the radiance $2$-form $\Lambda$.}
  \label{fig:two_form_pullback:geometry}
\end{figure*}

A differential $2$-form is an object that is naturally integrated over a surface $\mathcal{S} \subset \R^3$.
As discussed for example in the supplementary material, this requires to pull back the form to a chart of the surface.
For the radiance form $\Lambda$ we thus have to determine
\begin{subequations}
\begin{align}
  \int_{\mathcal{S}} \Lambda
  &= \int_{\varphi(\mathcal{S})} \left( \varphi^{-1} \right)^* \! \Lambda
  \\
  &= \int_{\varphi(\mathcal{S})} \Lambda \left( \dpp{}{u} , \dpp{}{v} \right) du \, dv
  \label{eq:radiance:pullback:def}
\end{align}
\end{subequations}
where the local basis vectors $\partial / \partial u$ and  $\partial / \partial v$ for the tangent space $T \mathcal{S}$ are given by
\begin{subequations}
\begin{align}
  \partial / \partial u = U^1 e_x + U^2 e_y + U^3 e_z
  \\
  \partial / \partial v = V^1 e_x + V^2 e_y + V^3 e_z
\end{align}
\end{subequations}
and $(U,\varphi)$ is a chart for $\mathcal{S}$, and we assume without loss of generality that all of $\mathcal{S}$ is covered by $(U,\varphi)$, see Fig.~\ref{fig:two_form_pullback:geometry}.
The pullback of $\Lambda$ is most conveniently computed using Eq.~\ref{eq:radiance:pullback:def} which, by linearity, can be determined term by term.
With Eq.~\ref{eq:radiance:modern:2}, we have for example for the first term
\begin{align}
  \left( L(q,\bar{p},\nu) \, \bar{p}_1 \, dy \wedge dz \right) \! \left( \dpp{}{u} , \dpp{}{v} \right)
  \label{eq:radiance:pullback:2}
\end{align}
where we also omitted the exterior basis functions unrelated to the surface integral.
Again using linearity, it is sufficient to consider the basis function for the pairing and treat $L(q,\bar{p},\nu) \, \bar{p}_1$ as a scalar factor.
Eq.~\ref{eq:radiance:pullback:2} can be simplified using
\begin{align}
  (\alpha \wedge \beta)(v_1 , v_2) = \alpha(v_1) \, \beta(v_2) - \alpha(v_2) \, \beta(v_1)
  \label{eq:wedge_product:2_form}
\end{align}
where $\alpha,\beta$ are arbitrary $1$-forms and $v_1,v_2$ are arbitrary vectors.\footnote{Cf. for example~\parencite[p. 67]{Frankel2003}.}
With Eq.~\ref{eq:wedge_product:2_form} we obtain
\begin{align*}
  ( dy \wedge dz ) \! \left( \dpp{}{u} , \dpp{}{v} \right)
  =& dy\left( U^1 e_x + U^2 e_y + U^3 e_3 \right) dz \left( V^1 e_x + V^2 e_2 + V^3 e_3 \right)
  \\
  - & dy\left( V^1 e_x + V^2 e_y + V^3 e_z \right) dz \left( U^1 e_x + U^2 e_y + U^3 e_z \right)
\end{align*}
and using the biorthogonality of the vector and covector basis functions this simplifies to
\begin{subequations}
\begin{align}
  ( dy \wedge dz ) \! \left( \dpp{}{u} , \dpp{}{v} \right)
  &= U^2 dy(e_y) \, V^3 dz(e_z) - V^2 dy(e_y) \, U^3 dz(e_z)
  \\
  &= U^2 V^3 - U^3 V^2 .
\end{align}%
  \label{eq:cosine:cross_product:1}%
\end{subequations}%
Similarly, for the other two $2$-form basis vectors we obtain
\begin{subequations}
\begin{align}
  ( dz & \wedge dx ) \! \left( \dpp{}{u} , \dpp{}{v} \right)
  = U^3 V^1 - U^1 V^3
  \label{eq:cosine:cross_product:2}
  \\
  ( dx & \wedge dy ) \! \left( \dpp{}{u} , \dpp{}{v} \right)
  = U^1 V^2 - U^2 V^1 .
  \label{eq:cosine:cross_product:3}
\end{align}
\end{subequations}
But the expressions in Eq.~\ref{eq:cosine:cross_product:1}--Eq.~\ref{eq:cosine:cross_product:3} are easily identified as the component form of the cross product.
Since $\partial / \partial u$ and $\partial / \partial v$ are the local basis vectors for the tangent space $T \mathcal{S}$, the pairing with the $2$-form basis functions thus yields the components of the local surface normal, that is
\begin{align}
  \vec{n}
  =
  \left( \! \! \begin{array}{c}
  n_1 \\ n_2 \\ n_3
  \end{array} \! \! \right)
  =
  \left( \! \begin{array}{c}
  U^2 V^3 - U^2 V^3 \\
  U^3 V^1 - U^3 V^1 \\
  U^1 V^2 - U^2 V^2
  \end{array} \! \right) .
\end{align}
With Eq.~\ref{eq:radiance:pullback:2}, the integrand in Eq.~\ref{eq:radiance:pullback:def} can now be written as
\begin{align}
  ( \varphi^{-1})^* \! \Lambda
 & = L(q,\bar{p},\nu)
 \left( \bar{p}_1 n_1 + \bar{p}_2 n_2 + \bar{p}_3 n_3 \right) \, du \, dv .
 \label{eq:radiance:cosine:final:1}
\end{align}
By writing the normal as $\vec{n} = \Vert \vec{n} \Vert \bar{n}$, where $\bar{n} = (\bar{n}_1 , \bar{n}_2 , \bar{n}_3 )$ is a unit vector, and with $\bar{p} \cdot \bar{n} = \bar{p}_1 \bar{n}_1 + \bar{p}_2 \bar{n}_2 + \bar{p}_3 \bar{n}_3$ we obtain
\begin{align}
  ( \varphi^{-1})^* \! \Lambda
  & = L(q,\bar{p},\nu) (\bar{p} \cdot \bar{n} )  \Vert \vec{n} \Vert \, du \, dv .
 \intertext{But by definition $dA = \Vert \vec{n} \Vert \, du \, dv$ so that}
  ( \varphi^{-1})^* \! \Lambda
  & = L(q,\bar{p},\nu) (\bar{p} \cdot \bar{n}) \, dA .
\end{align}
The vectors $\bar{p}$ and $\bar{n}$ are of unit length so that $\bar{p} \cdot \bar{n} = \cos{\theta}$ where $\theta$ is the angle between the vectors.
We thus have for the infinitesimal area flux through the surface $\mathcal{S}$ in direction $\bar{p}$ that
\begin{align}
   \Lambda &\left( \dpp{}{u} , \dpp{}{v} \right) du \, dv
   =  L(q,\bar{p},\nu) \cos{\theta} \, dA \, d\bar{p} \, d\nu  .
   \label{eq:radiance:cosine_term:final}
\end{align}
The cosine term in Eq.~\ref{eq:radiance:cosine_term:final} appeared rigorously from our formulation of radiance as a differential $2$-form without the need for postulates or heuristic arguments.
It should also be noted that the dot product from which the cosine term was obtained was not derived from the Riemannian structure on $\R^3$ but solely resulted from the pullback.

Returning briefly to Eq.~\ref{eq:radiance:cosine:final:1}, by linearity the equation can be written as
\begin{align}
  L(q,\bar{p},\nu) & \, (\bar{p} \cdot \vec{n}) \, du \, dv \, d\bar{p} \, d\nu
  = \vec{n} \cdot \left( L(q,\bar{p},\nu) \, \bar{p} \, d\bar{p} \, d\nu \right)  du \, dv .
\end{align}
With the definition of vector irradiance in Eq.~\ref{eq:vector_irradiance}, we therefore have that the effective surface flux, the irradiance $I(q)$, at a point $q \in \R^3$ is
\begin{align}
  I(q) dt
  = \bar{n} \cdot \left( \int_{\Omega_q} L(q,\bar{p},\nu) \, d\bar{p} \right) dA
  = \bar{n} \cdot J(q) \, dA
\end{align}
where $\Omega \subseteq S^2$ is now a finite solid angle, for example the solid angle subtended by a light source.
Note that the possibility of computing the infinitesimal surface flux using a dot product was the original motivation for introducing vector irradiance.\footcite{Gershun1936}

%%%%%%%%%%%%%%%%%%%%%%%%%%%%%%%%%%%%%%%%%%%%%%%%%%%%%%%%%%%%%%%%%%%%%%%%%%%%%%%%
\subsection{The Area Formulation of Light Transport}
\label{sec:radiance:area}

In computer graphics, usually the parametrization of radiance over solid angle is employed.
However, as was emphasized for example by Veach,\footcite{Veach_phd} the area formulation is equally important.\footnote{The area formulation goes again back to Lambert.\footcite{Lambert1760} It seems however that Lambert obtained the equation only for emitters satisfying the Lambert emission law.}
Historically it was in fact this parametrization that was used most frequently since it enables to obtain closed form solutions for the incident light energy.
%A general solution is however not possible, although for two polygons a  solution was obtained by Schr{\"o}der and Hanrahan~\cite{Schroder1993}.

\begin{figure}[t]
  \begin{center}
    \includegraphics[trim = 65mm 105mm 60mm 137mm, clip, scale=0.8]{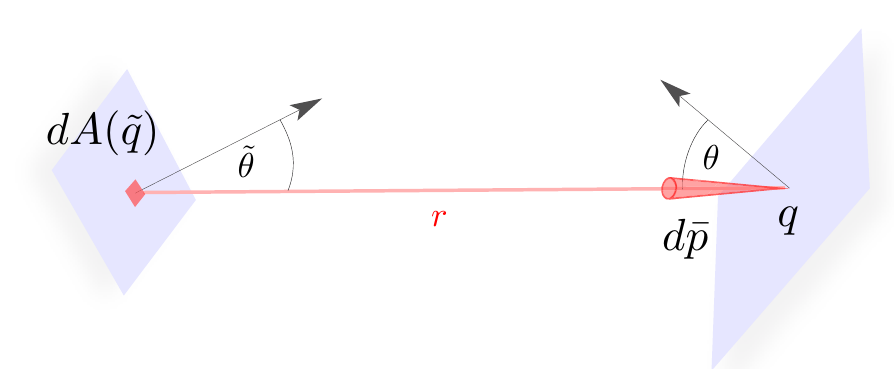}
\end{center}
  \caption{Geometry of the area formulation.}
  \label{fig:area_form:geometry}
  \vspace{-0.2in}
\end{figure}

Central to the area formulation is the change of variables\footnote{\parencite[p. 24]{Dutre2006},~\parencite[p. 292]{pbrt2}.}
\begin{align}
  d\bar{p} = \frac{\cos{\tilde{\theta}}}{r^2} dA(\tilde{q})
  \label{eq:area:classical}
\end{align}
relating the solid angle $d\bar{p}$ to an infinitesimal area $dA(\tilde{q})$ along the flow emanating from $d\bar{p}$, see Fig.~\ref{fig:area_form:geometry}.
To derive Eq.~\ref{eq:area:classical}, we will employ the $2$-form
\begin{align}
  \beta
  = \frac{x \, dy \wedge dz + y \, dz \wedge dx + z \, dx \wedge dy }{\left( x^2 + y^2 + z^2 \right)^{3/2}}
\end{align}
where $\vec{p} = (x,y,z)$ and one should think of $\beta \in \Omega^2(\R^3 \! \setminus \{ 0 \})$ as being centered at $q \in \R^3$ and ``living'' in the cotangent space $T_q^* \R^3 \cong \R^3$.\footnote{Our derivation follows~\parencite[Exercise 5-31]{Spivak1971}.}
%The change of variables in Eq.~\ref{eq:area:classical} will be obtained from $\beta$ by showing that its pullback onto an arbitrary surface equals the right hand side of the equation, and that the integral over the surface equals the solid angle subtended by the surface.

\begin{remark}
  {\itshape
  The $2$-form $\beta \in \Omega^2( \R^3 \! \setminus \! \{ 0 \})$ is formed by
  \begin{align}
    \beta = \frac{1}{\Vert \vec{p} \Vert^2} dA(\vec{p})
    \label{eq:beta:explanation:1}
  \end{align}
  where $\vec{p}$ has the natural interpretation as the direction of propagation.
  Equating $\vec{p} = \vec{n} = (n_1,n_2,n_3)$, the surface $2$-form $dA \in \Omega^3(\R^3 \! \setminus \! \{ 0 \} )$ is given by\footcite[Theorem 5-6]{Spivak1971}
  \begin{align}
    dA(\vec{n}) = \frac{1}{\Vert \vec{n} \Vert} \, n_1 \, dy \wedge dz + n_2 \, dz \wedge dx + n_3 \, dx \wedge dy .
    \label{eq:beta:explanation:2}
  \end{align}
  The factor of $1 / \Vert \vec{n} \Vert^2$ in Eq.~\ref{eq:beta:explanation:1} thus ensures that the integral of $\beta$ over $S^2(r)$, the sphere with radius $r$, is always $4 \pi$.
  This can be interpreted as always considering an equal size surface element of $S^2(r)$.
  }
\end{remark}

We will begin by showing that $\beta$ is a closed differential form, that is $\dd \beta = 0$.
By linearity, we can compute the exterior derivative term by term, that is
\begin{align}
  \dd \beta \!
  = \! \dd \! \left( \!\! \frac{x}{\Vert \vec{p} \Vert^{3}} dy \wedge dz \! \right)
  + \dd \! \left( \!\! \frac{y}{\Vert \vec{p} \Vert^{3}} dz \wedge dx \! \right)
  + \dd \! \left( \!\! \frac{z}{\Vert \vec{p} \Vert^{3}} dx \wedge dy \! \right)
\end{align}
where $\Vert \vec{p} \Vert^{3} = ( x^2 + y^2 + z^2 )^{3/2}$.
For the first term of the exterior derivative we obtain
\begin{align}
  \dd \left( \frac{x}{\Vert \vec{p} \Vert^{3}} \right) dy \wedge dz
  = \dpp{}{x} \frac{x}{\left( x^2 + y^2 + z^2 \right)^{3}} dx \wedge dy \wedge dz
\end{align}
where the remaining partial derivatives vanish by the anti-symmetry of differential forms.
Using the product rule and after collecting terms, we then have
\begin{subequations}
\begin{align}
  \dd \left( \frac{x}{\Vert \vec{p} \Vert^{3}} \right) dy \wedge dz
  &= \frac{-2y + y^2 +z^2}{\left( x^2 + y^2 + z^2 \right)^{5/2}} dx \wedge dy \wedge dz.
  \intertext{Analogously, for the other two terms of $\dd \beta$ we obtain}
  \dd \left( \frac{y}{\Vert \vec{p} \Vert^{3}} \right) dz \wedge dx
  &= \frac{y - 2 y^2 +z^2}{\left( x^2 + y^2 + z^2 \right)^{5/2}} dx \wedge dy \wedge dz
  \\
  \dd \left( \frac{z}{\Vert \vec{p} \Vert^{3}} \right) dx \wedge dy
  &= \frac{y + y^2 - 2 z^2}{\left( x^2 + y^2 + z^2 \right)^{5/2}} dx \wedge dy \wedge dz .
\end{align}
\end{subequations}
With the above, it is immediately apparent that we have
\begin{align}
  \dd \beta
  = \frac{0}{\left( x^2 + y^2 + z^2 \right)^{5/2}} dx \wedge dy \wedge dz
  = 0
\end{align}
and hence $\beta \in \Omega^2(\R^3 \! \setminus \! \{ 0 \} )$ is indeed closed.

In the following, we will also need the pullback of $\beta \in \Omega^2(\R^3 \! \setminus \! \{ 0 \} )$ onto an arbitrary surface $\mathcal{S} \in \R^3$.
Following an argument analogous to those in Sec.~\ref{sec:radiance:cosine}, we obtain for the pullback
\begin{align}
  \left(\varphi^{-1} \right)^* \beta
  = \frac{\vec{p} \cdot \vec{n}}{\Vert \vec{p} \Vert^3} du \, dv
\end{align}
where $\vec{n}$ is again the local surface normal.
Writing $\vec{p} = \Vert \vec{p} \Vert \bar{p}$ and $\vec{n} = \Vert \vec{n} \Vert \bar{n}$, where $\bar{p}$ and $\bar{n}$ are unit vectors, we obtain
\begin{align}
  \left(\varphi^{-1} \right)^* \beta
  = \frac{\Vert \vec{p} \Vert \, \Vert \vec{n} \Vert (\bar{p} \cdot \bar{n})}{\Vert \vec{p} \Vert^3}  du \, dv
  = \frac{ (\bar{p} \cdot \bar{n}) }{\Vert \vec{p} \Vert^2} dA
  \label{eq:radiance:area:4b}
\end{align}
which is the right hand side of Eq.~\ref{eq:area:classical}.
Note that when the surface $\mathcal{S}$ coincides with a subset of the sphere $S^2$ so that the cosine term is unity and $\Vert \vec{p} \Vert = 1$, then we have $(\varphi^{-1})^* \beta = d\bar{p}$, that is, the pullback of $\beta$ yields the solid angle measure.

\begin{figure}
  \setlength{\abovecaptionskip}{-15pt}
  \begin{center}
  \includegraphics[trim = 60mm 175mm 50mm 18mm, clip, scale=1.0]{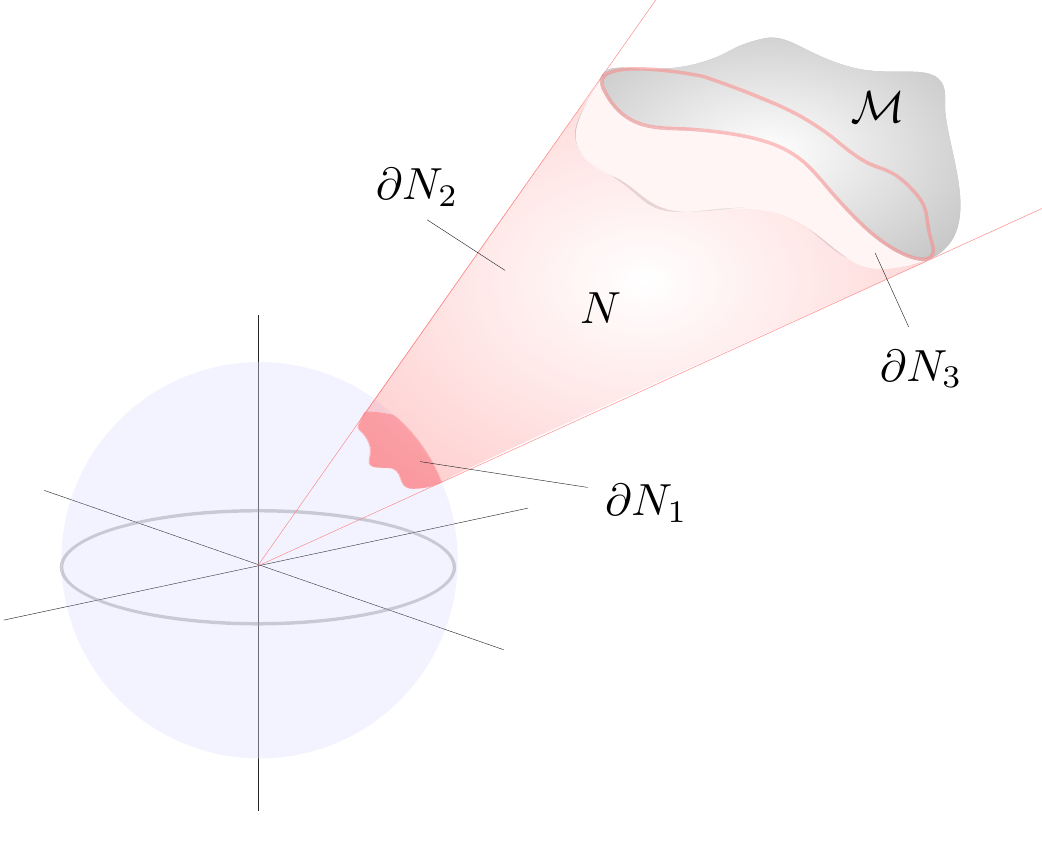}
  \end{center}
  \caption{Geometry employed in the proof of the area formulation of light transport.}
  \label{fig:area_cone}
  \vspace{-0.2in}
\end{figure}

With $\dd \beta = 0$ and Eq.~\ref{eq:radiance:area:4b}, we can finally establish the change of variables in Eq.~\ref{eq:area:classical}.
For this, consider a volume $N \subset \R^3$ as shown in Fig.~\ref{fig:area_cone}, bounded by the surface of a manifold $\M$ and a subset of the sphere.
Since $\beta \in \Omega^2(\R^3 \! \setminus \! \{ 0 \} )$ is closed, we have by Stokes theorem for differential forms that
\begin{subequations}
\begin{align}
  0  &= \int_N \dd \beta = \int_{\partial N} \beta
  \intertext{where $\partial N$ is the boundary of $N$. Since $\partial N$ consists of three parts, we can write}
  0 &= \int_{\partial N_1} \beta + \int_{\partial N_2} \beta +
\int_{\partial N_3} \beta .
\intertext{The integral over $\partial N_2$ vanishes since the radial vector $\vec{p}$ and the normal $\vec{n}$ of $\partial N_2$ are orthogonal. We hence have}
  0 &= \int_{\partial N_1} \beta + \int_{\partial N_3} \beta .
\end{align}
\end{subequations}
By our previous results, the first term equals the solid angle subtended by $\M$ while the second term is the integral over Eq.~\ref{eq:radiance:area:4b}.
Since $\M$ was arbitrary, the result must also hold infinitesimally, that is,
\begin{align}
  d\bar{p} = \frac{\cos{\theta}}{\Vert \vec{p} \Vert^2} dA .
\end{align}

\section{Discussion}

Radiance is the fundamental concept in radiometry, all other radiometric quantities being derived from it by integration.
Nonetheless, its mathematical formulation still employs the classical infinitesimals that were introduced by Lambert.
Moreover, while intrinsically related to measurements, it is today also employed to model the transport of light intensity.
These inadequacies lead to numerous theoretical and practical problems, hampering the development of more effective computational techniques.
In this paper, we provide a first step toward a modern theory of light transport by addressing the question of measurements and introducing a modern formulation of radiance.
In contrast to classical radiance, our formulation as a differential form is vector-valued and becomes scalar only when paired to a surface, showing that radiance is the area flux through a yet undetermined surface.
%, introducing for example the need to employ postulates or heuristic arguments to establish the cosine term and the area formulation,

We demonstrated the utility of our geometric formulation of radiance by rigorously deriving the cosine term and the area formulation, results that previously were either postulated or justified using heuristic geometric arguments.
Our derivation also demonstrates that radiance falls off quadratically for measurements on concentric spheres when the area is kept constant, and that the classical results that ``radiance is constant along a ray'' has to be reconsidered, since radiance is not intrinsically a transported quantity, cf. the appendix.

Readers familiar with vector calculus will have noticed that the result in Sec.~\ref{sec:radiance:cosine} could have been obtained using classical means starting from a vector formulation of radiance, cf. Eq.~\ref{eq:radiance:modern:components}.
However, to the authors' knowledge, this is no onger the case for the derivation of the area formulation which requires a formulation of radiance as a differential form and the power of exterior calculus---also providing much added insight.
Additionally, it is also the radiance $2$-form, and not a vector, that can be derived from Maxwell's equations by considering measurements within the modern, geometric theory of light transport, as was developed in the first authors Ph.D. disseration\footcite{Lessig_phd} and will be presented in a forthcoming publication.

%The formulation of radiance as a differential form we have presented provides the first step toward a modern theory of light transport that will be developed in~\cite{Lessig2012x}, a formulation that will provide a contemporary perspective for the propagation of electromagnetic energy in macroscopic environments, and which will pave the way for many novel and exciting applications in computer graphics and the many other fields where the theory is employed.

%%%%%%%%%%%%%%%%%%%%%%%%%%%%%%%%%%%%%%%%%%%%%%%%%%%%%%%%%%%%%%%%%%%%%%%%%%%%%%%%
\subsection*{Acknowledgements}
We would like to thank Tyler de Witt and Jos Stam for helpful discussions on the nature of radiance and its formulation using exterior calculus.
Financial support by NSERC, GRAND, and NSF grant CCF-1011944 is gratefully acknowledged.

%%%%%%%%%%%%%%%%%%%%%%%%%%%%%%%%%%%%%%%%%%%%%%%%%%%%%%%%%%%%%%%%%%%%%%%%%%%%%%%%
%%%%%%%%%%%%%%%%%%%%%%%%%%%%%%%%%%%%%%%%%%%%%%%%%%%%%%%%%%%%%%%%%%%%%%%%%%%%%%%%
\printbibliography[maxnames=20]

%%%%%%%%%%%%%%%%%%%%%%%%%%%%%%%%%%%%%%%%%%%%%%%%%%%%%%%%%%%%%%%%%%%%%%%%%%%%%%%%
\newpage
\appendix
%\begin{figure}[!tp]
%    \frame{
%    \begin{minipage}{1.0\columnwidth}
%      \setlength{\leftskip}{0.05in}
%      \setlength{\rightskip}{0.05in}
%      \vspace{0.05in}
%      \tighteq
%{\bfseries The transport of density.}
\section{The Transport of a Density}
If radiance is only meaningful for measurements, how does one then describe the transport of light energy?
To understand this, we will consider the transport of mass in fluid dynamics.
Infinitesimally, the mass is described by the mass density $\rho = \sigma(q) \, dq \in \Den{(Q)}$, an element in $\Den(Q)$ which, roughly speaking, is equivalent to the space of volume forms $\Omega^3(Q)$ on $Q \subset \R^3$.
The transport of the density in a fluid with velocity vector field $\vec{u} \in \mathfrak{X}(Q)$ is then given by 
%the continuity equation
\begin{align*}
  \dot{\rho} + \Lie_{\vec{u}} \rho = 0 ,
\end{align*}
where $\Lie_{\vec{u}} \rho$ is the Lie derivative, or, in classical notation,
\begin{align*}
  \dot{\rho} + \nabla \cdot (\sigma \, \vec{u}) = 0 .
\end{align*}

\includegraphics[trim = 35mm 70mm 30mm 70mm, clip, scale=0.45]{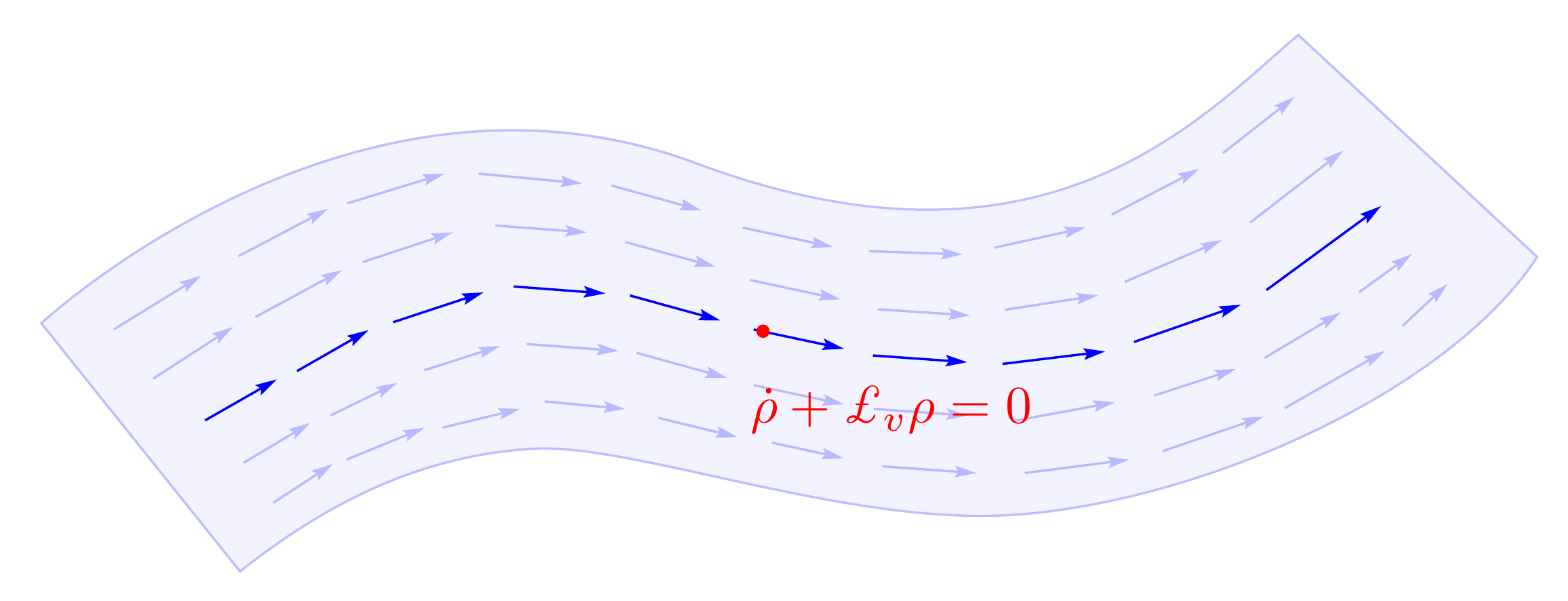}

Let us now determine the mass flow through an infinitesimal surface submerged in the fluid.
Using the transport theorem, a central result in continuum mechanics\footcite[Theorem 8.1.12]{Marsden2004}, we obtain that the flow through a finite surface $S \subset \R^3$ is
\begin{align*}
  \int_S (\sigma \, \vec{u}) \cdot \bar{n} \, dA
\end{align*}
where $\vec{n}$ is the surface normal.
Infinitesimally, we thus have
\begin{align*}
  \sigma \, \vec{u} \cdot \bar{n} \, dA
\end{align*}
which up to a factor that depends on the local magnitude of the fluid velocity vector field is equivalent to
\begin{align*}
  \sigma \, \bar{u} \cdot \vec{n} \, dA = \sigma \, \cos{\theta} \, dA ,
\end{align*}
where $\bar{u}$ is the normalized velocity vector and $\theta = \angle( \vec{u},\vec{n})$.
When expressed using exterior calculus, the integrand $\rho \vec{u}$ is a $2$-form, and given by $\rho(q) \, dA_u$ where $dA_u = u_1 \, dq^2 \wedge dq^3 + u_2 \, dq^3 \wedge dq^1 + u_3 \, dq^1 \wedge dq^2$.

\includegraphics[trim = 35mm 70mm 30mm 70mm, clip, scale=0.45]{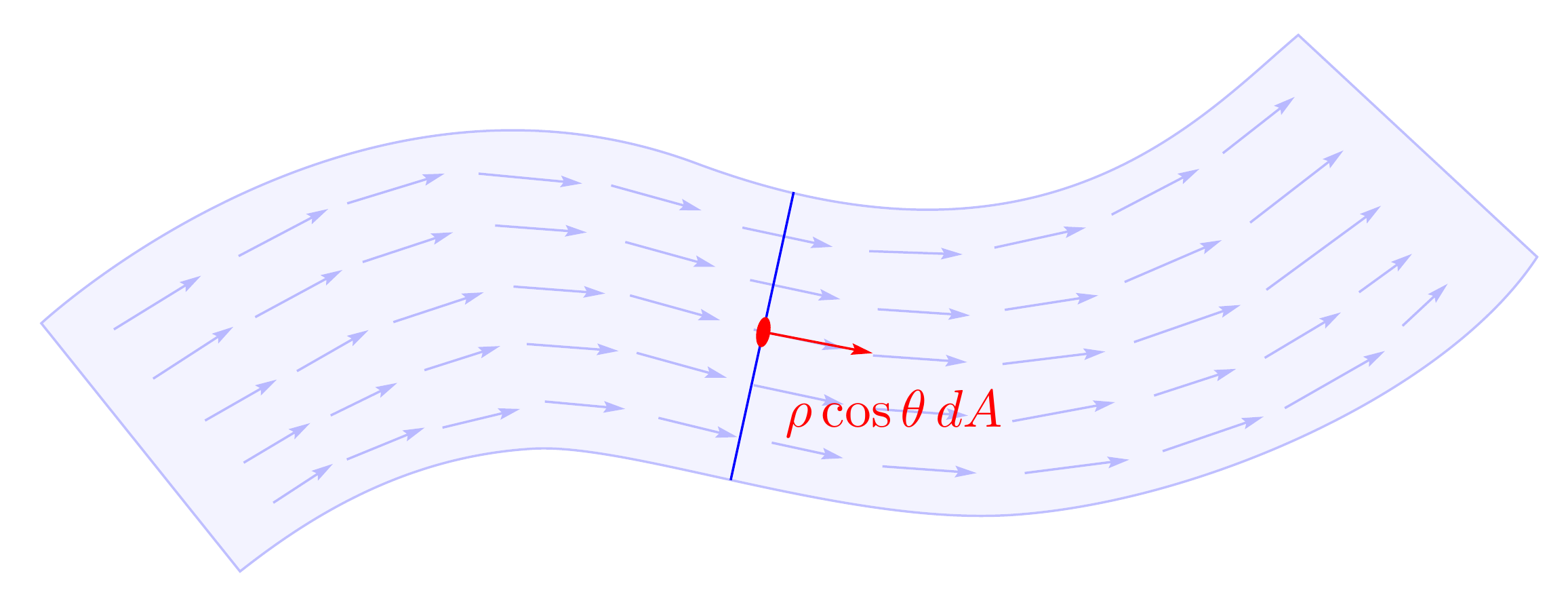}

As will be shown in a forthcoming publication, the quantity that is transported in light transport theory is the phase space light energy density, or, short, light energy density,
\begin{align}
  \ell = L(q,p) \, dq \, dp \in \Den{( T^*Q )}
\end{align}
which is a density on the six-dimensional phase space $T^*Q \cong Q \times \R^3$ for $Q \subset \R^3$.
The light energy density $\ell \in \Den(T^*Q)$ is the light transport equivalent of the mass density $\rho \in \Den(Q)$, and the expression $\sigma(q) \cos{\theta} \, dA$ is the fluid equivalent of classical radiance $L(x,\omega) \cos{\theta} \, d\omega \, dA$, describing measurements through an infinitesimal surface.

%\vspace{0.05in}
%    \end{minipage}
%    }
%\end{figure}

\end{document}